\documentclass[superscriptaddress,aps,journal=prl,twocolumn]{revtex4-2}
\usepackage{graphicx}
\usepackage{epstopdf}
\usepackage{bm,amsmath,amssymb} % for math
\usepackage{color, soul} % for highlight
\usepackage{dcolumn}   % needed for some tables
\usepackage{multirow}

\begin{document}

\
\title{On-demand quantum spin Hall insulators controlled by two-dimensional ferroelectricity}% Force line breaks with \\

\author{Jiawei Huang}
\affiliation{School of Science, Westlake University, Hangzhou, Zhejiang 310024, China}
\author{Xu Duan}
\affiliation{School of Science, Westlake University, Hangzhou, Zhejiang 310024, China}
\author{Sunam Jeon}
\affiliation{Department of Energy Science,
Sungkyunkwan University, Suwon 16419, Korea}
\author{Youngkuk Kim}
\affiliation{Department of Physics
Sungkyunkwan University, Suwon 16419, Korea}
\author{Jian Zhou}
\affiliation{Center for Alloy Innovation and Design, State Key Laboratory for Mechanical Behavior of Materials, Xi’an Jiaotong University, Xi’an 710049, China}
\author{Jian Li}
\email{lijian@westlake.edu.cn}
\affiliation{School of Science, Westlake University, Hangzhou, Zhejiang 310024, China}
\affiliation{Institute of Natural Sciences, Westlake Institute for Advanced Study, Hangzhou, Zhejiang 310024, China}
\affiliation{Key Laboratory for Quantum Materials of Zhejiang Province, Hangzhou Zhejiang 310024, China}
\author{Shi Liu}
\email{liushi@westlake.edu.cn}
\affiliation{School of Science, Westlake University, Hangzhou, Zhejiang 310024, China}
\affiliation{Institute of Natural Sciences, Westlake Institute for Advanced Study, Hangzhou, Zhejiang 310024, China}
\affiliation{Key Laboratory for Quantum Materials of Zhejiang Province, Hangzhou Zhejiang 310024, China}

\date{\today}

\begin{abstract}{ 
The coexistence of ferroelectric and topological orders in two-dimensional (2D) atomic crystals allows non-volatile and switchable quantum spin Hall states. Here we offer a general design principle for 2D bilayer heterostructures that can host ferroelectricity and nontrivial band topology simultaneously using only topologically trivial building blocks. The built-in electric field arising from the out-of-plane polarization across the heterostrucuture enables a robust control of the band gap size and band inversion strength, which can be utilized to manipulate topological phase transitions. Using first-principles calculations, we demonstrate a series of bilayer heterostructures are 2D ferroelectric topological insulators (2DFETIs) { characterized with a direct coupling between band topology and polarization state}. We propose a few 2DFETI-based quantum electronics including domain-wall quantum circuits and topological memristor. 
}
\end{abstract}

\maketitle
\newpage
Band topology and ferrroelectricity, two extensively studied  properties of bulk insulators representing two distinct ``ordered states", can manifest themselves in two dimensions. Graphene as the first discovered two-dimensional (2D) material~\cite{Novoselov04p5696} is also the first predicted 2D topological insulators (TIs) characterized by counter-propagating edge currents with opposite spin polarization and an insulating interior~\cite{Kane05p226801}. A 2D TI is also called a quantum spin Hall (QSH) insulator for its quantized edge conductance ($2e^2/h$ where $e$ is the elementary charge and $h$ is Planck’s constant). Finding 2D TIs with large band gaps for room-temperature applications remains an actively pursued goal~\cite{Qian14p1344,Kou17p1905}. 

Ferroelectrics (FEs) with inversion symmetry breaking often exhibit a strong size effect that the spontaneous polarization diminishes with reduced dimensionality due to the depolarization field from the incomplete screening of surface charges~\cite{Batra73p3257}. More recently, facilitated by first-principles calculations based on density functional theory (DFT), a range of 2D FEs were discovered followed by confirming experiments~\cite{Belianinov15p3808,Zhou17p5508, Chang16p274, Yuan19p1775,Xiao20p1028}. Specifically, $\alpha$-In$_2$Se$_3$ exhibits both out-of-plane and in-plane electric polarization~\cite{Ding17p14956, Zhou17p5508}, a feature beneficial for practical device applications and high-density integrations. As the surfaces of 2D materials do not suffer from dangling bonds, it is feasible to stack different 2D sheets to construct van der Waals (vdW) heterostructures in a precisely controlled layering sequence less impacted by the lattice mismatch~\cite{Duong17p11803}. 

A 2D material system with both ferroelectric and topological orders, referred to as a 2D ferroelectric topological insulator (2DFETI), remains rarely reported other than some chance discovery~\cite{Zhang20p785}. The coexistence of ferroelectricity and nontrivial band topology in general has to reconcile conflicting requirements for band gaps~\cite{Liu16p1663}; TIs are often narrow-gap semiconductors with a band gap determined by the strength of spin-orbit coupling (SOC) whereas archetypal FEs such as transition metal perovskites are mostly wide-band-gap insulators with the gap size dictated by the electronegativity difference between oxygen and transition metals. Unlike bulk FEs in 3D, many 2D FEs are semiconductors with moderate band gaps~\cite{Ding17p14956}, thus being better suited for the coexistence of the topological order. { However, a mere coexisting of these two ordered states does not guaranty a strong coupling between topological and polarization states.}
\begin{figure}[t]
\centering
\includegraphics[scale=1]{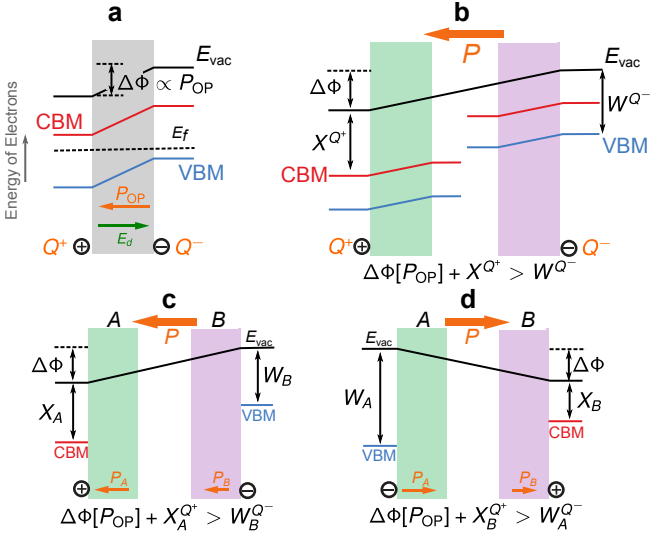}
 \caption{Design principle for two-dimensional ferroelectric topological insulator. Band bending in ({\bf a}) 2D ferroelectric and ({\bf b}) bilayer heterostructure consisted of two different 2D ferroelectrics. { The solid line represents the energy of electrons.} The band inversion can be controlled by the switching of ferroelectric polarization. ({\bf c}) Potential topological insulator phase due to band inversion. ({\bf d}) Trivial insulator phase with uninverted bands. }
  \label{model}
 \end{figure}

Here we propose a design principle for the realization of 2DFETIs in bilayer heterostructures comprising only trivial 2D FEs. { The ability to create nontrivial quantum materials using trivial building blocks broadens the materials design space. Moreover, the band topology is directly coupled to the polarization state: the nonvolatile QSH states that can be fully switched on and off via voltage.} The key requirement for the constituent 2D FEs is the presence of an out-of-plane polarization ($P_{\rm OP}$). { For a free-standing monolayer,} the polarization bound charges on the two surfaces create a depolarization field ($E_d$) that runs against the polarization. As a result, the valence band maximum (VBM) is located on the negatively-charged ({ $Q^-$}) surface while the conduction band minimum (CBM) is on the positively-charged ( $Q^+$) surface (Fig.\,\ref{model}a). It reflects the tendency of the system to generate free carriers needed for bound charge compensation: {the screening of  $Q^-$ ($Q^+$) surface needs mobile holes (electrons) that can be generated by the crossing of VBM (CBM) over the Fermi level ($E_F$). Another heuristic way to understand such band bending is that electrons close to the $Q^-$ surface are at a high energy level (thus being at VBM) because of the Coulomb repulsion. The band diagram of a 2D FE with out-of-plane polarization resembles that of an unbiased $p$-$n$ junction, and $E_d$ across the monolayer is similar to the electric field confined to the depletion region around the junction interface (see Fig.\,S2 in Supplemental Materials).} Since the potential step ($\Delta \Phi$) scales with $P_{\rm OP}$, one might expect the crossover of VBM and CBM given a sufficiently large $P_{\rm OP}$. However, such band inversion is unlikely to happen in a monolayer as the system would become metallic at the crossover, providing more free carriers for surface charge compensation thus reducing $E_d$ and band bending. Therefore, a 2D FE by itself has a tendency to avoid the band inversion by maintaining an optimal trade-off between imperfect screening and a minimal necessary band bending~\cite{Sluka12p748}. 

In contrast, a bilayer heterostructure made of two different 2D FEs allows for a band inversion process when the following condition is satisfied
\begin{equation}
\Delta \Phi [P_{\rm OP}] + X^{Q^+} > W^{Q^-},
\label{req1}
\end{equation}
where  $X^{Q^+}$  is the electron affinity of the $Q^+$ surface and $W^{Q^-}$ is the work function of the $Q^-$ surface of the heterosrtructure, respectively. The superscripts ($Q^+$ and $Q^-$) are simply used to reflect the direction of out-of-plane polarization. It is evident from Fig.\,\ref{model}b that $E_{\rm VBM}=-W^{Q^-}$ and $E_{\rm CBM}=-(\Delta \Phi + X^{Q^+} )$ relative to the vacuum level, and Eq.\,\ref{req1} naturally leads to $ E_{\rm CBM}<E_{\rm VBM}$ and thus a band inversion. {According to the celebrated Neumann-Wigner theorem~\cite{Neumann27p467}, the presence of crossing between two bands often demands a symmetry-related protection (e.g., mirror symmetry). For a generic system without special crystalline symmetries, the SOC will then lead to a double group system where a gap generically opens between valence and conduction bands, likely resulting in a QSH state.} Because $X^{Q^+}$ and $W^{Q^-}$ are coming from two different materials, a 2DFETI can be realized by selecting one layer with large $X$ and another layer with small $W$.   

\begin{figure}[b]
\centering
\includegraphics[scale=1]{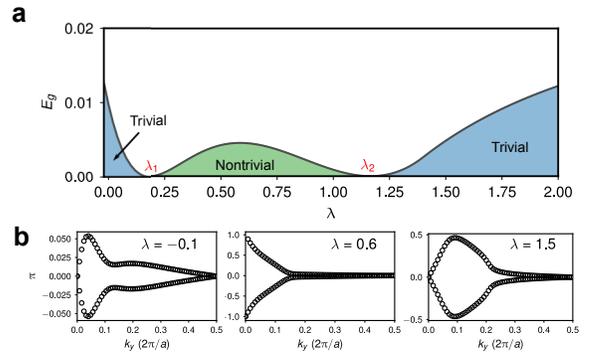}
 \caption{Successive trivial-nontrivial-trivial phase transitions driven by the strength ($\lambda$) of band inversion. ({\bf a}) Numerically obtained band gap as a function of $\lambda$. ({\bf b}) Wilson loop calculations for three different phases.}
  \label{phase}
 \end{figure}
  
Furthermore, by choosing a pair of 2D FEs (labeled as $A$ and $B$ respectively) satisfying
\begin{equation}
\Delta \Phi [P_{\rm OP}] + X_A^{{Q^+}} > W_B^{{Q^-}}; \Delta \Phi [P_{\rm OP}] + X_B^{{Q^+}} < W_A^{{Q^-}}, 
\label{ti2ni}
\end{equation}
we can ensure that the configuration with $P$ pointing from $B$ to $A$ has band inversion (Fig.\,\ref{model}c) while the other configuration with $P$ pointing from $A$ to $B$ is a normal insulator (Fig.\,\ref{model}d), creating a pair of topologically different, non-volatile states in the absence of external electric fields. This can be accomplished by choosing $A$ with large $X$ and $W$ whereas $B$ with small $W$ and $X$. 

 \begin{figure*}[t]
\centering
\includegraphics[scale=1]{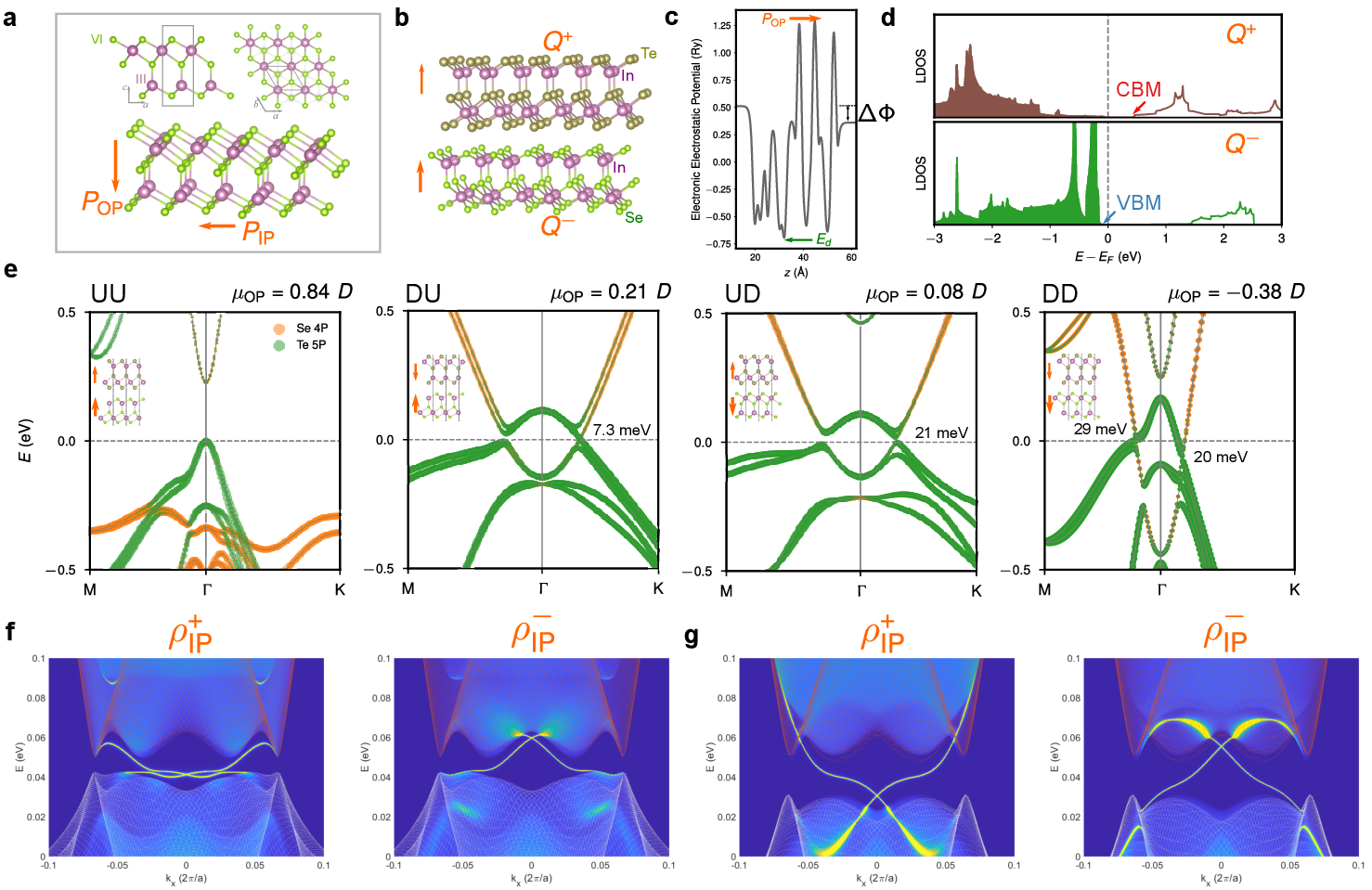}
 \caption{ Electronic structures of In$_2$Te$_3$/In$_2$Se$_3$ bilayer heterostructures. Atomic structures of ({\bf a}) III$_2$-VI$_3$ 2D ferroelectrics and ({\bf b}) In$_2$Te$_3$/In$_2$Se$_3$ bilayer hetrostructure of UU configuraiton. ({\bf c}) Electronic electrostatic potential of UU configuration and ({\bf f}) layer-resolved local density of states (LDOS) for $Q^+$ and $Q^-$ surfaces computed with DFT. The potential step across the bilayer and the locations of VBM and CBM are consistent with Fig.\,1d. ({\bf e}) Atomic orbital-resolved band structures of UU, DU, UD, and DD configurations. $\mu_{\rm OP}$ is the out-of-plane dipole. Spectral functions at the edges of ({\bf f}) DU and ({\bf g}) UD configurations. Because of the in-plane polarization $P_{\rm IP}$, the two edges acquire opposite bound charges ($\rho_{\rm IP}^+$ and $\rho_{\rm IP}^-$), leading to shifted Dirac cones relative to the Fermi level. }
  \label{UUDD}
 \end{figure*}  
 
{An advantage of the proposed 2DFETI is the ``on-demand" topological quantum phase transition.
In general, $X$ and $W$ are less sensitive to the change of polarization for a given material; the potential step $\Delta \Phi$ is almost singularly determined by $P_{\rm OP}$ and $E_d$. The band inversion strength ($\lambda \propto {E_d}$) can thus be tuned nearly continuously by applying external electric/stress fields. This implies the access to multiple electronic band configurations that are effectively characterized by the magnitude of $\lambda$ and can belong to distinct topological phases. Using a generic model (see details in Supplemental Materials),  we explore the evolution of the band topology with increasing value of $\lambda$ and identify successive phase transitions associated with two band inversion events, occurring at $\Gamma$-point and then non-high-symmetry points in the Brillouin zone. The two quantum phase transition points (band gap $E_g=0$, see Fig.\,\ref{phase}a) separate trivial, nontrivial, and trivial phases, respectively, as verified by Wilson loop calculations (Fig.\,\ref{phase}b).}

Using bilayer heterostructures made of III$_2$-VI$_3$-type 2D FEs as model systems, we demonstrate the feasibility of the design principle based on DFT calculations. A typical III$_2$-VI$_3$-type 2D FE comprises five atomic planes in the order of VI-III-VI-III-VI, in which the central layer is displaced relative to the top and bottom III-VI layers, resulting in both out-of-plane and in-plane polarization ($P_{\rm OP}$ and $P_{\rm IP}$, Fig.\,\ref{UUDD}a).
DFT calculations are performed using generalized gradient approximation of the Perdew-Burke-Ernzerhof (PBE) type as implemented in \texttt{QUANTUM ESPRESSO}~\cite{Giannozzi09p395502, Giannozzi17p465901} (see Supplemental Materials). To recognize the topological state of this system, we obtained the $\mathbb{Z}_2$ topological index by calculating the Wilson loop spectrum using the maximum localized Wannier functions tight-binding Hamiltonian constructed with Wannier90~\cite{Pizzi20p165902} interfaced with \texttt{QUANTUM ESPRESSO}.

We focus on the bilayer heterostructure made of monolayer In$_2$Te$_3$ and In$_2$Se$_3$, which has $W_{\rm{In_2Se_3}}=6.0$~eV, $X_{\rm{In_2Se_3}}=4.7$~eV, $W_{\rm{In_2Te_3}}=5.3$~eV, and $X_{\rm{In_2Te_3}}=3.8$~eV. { Noted that ${\rm{In_2Se_3}}$ has large $W$ and $X$ compared to ${\rm{In_2Te_3}}$, hinting at a possible material realization of Eq.\,\ref{ti2ni}.} The configuration with out-of-plane polarization pointing from In$_2$Se$_3$ ($Q^-$ surface) to In$_2$Te$_3$ ($Q^+$ surface) is denoted as UU (up-up,  Fig.\,\ref{UUDD}b).  { The computed electronic electrostatic potential across the UU bilayer (Fig.\,\ref{UUDD}c) is consistent with the design principle illustrated in Fig.\,\ref{model}d. The layer-resolved local density of states confirm that the VBM and CBM are located at $Q^-$ and $Q^+$ surfaces, respectively.}  According to Eq.\,\ref{ti2ni}, the UU configuration is a normal insulator as $ X_{\rm{In_2Te_3}}^{Q^+} \ll W_{\rm{In_2Se_3}}^{Q^-}$. In comparison, the configuration with a reversed polarization, denoted as DD (down-down), has In$_2$Se$_3$ being the $Q^+$ surface and In$_2$Te$_3$ being the $Q^-$ surface. Then the band inversion can occur if $\Delta \Phi + X_{\rm{In_2Se_3}}^{Q^+}> W_{\rm{In_2Te_3}}^{Q^-}$; the threshold value of $\Delta \Phi$ is just 0.6~eV, which is much more feasible. 

The hetero-bilayer also allows two configurations with intermediate values of $P_{\rm OP}$, denoted as UD and DU (up-down and down-up, respectively, illustrated in the insets of Fig.\,\ref{UUDD}e). Figure\,\ref{UUDD}e shows the atomic-orbital resolved band structures of In$_2$Te$_3$/In$_2$Se$_3$ of four different configurations including the effects of SOC. The computed band structures reveal a band inversion between Se-$4p$ and Te-$5p$ states in DU, UD, and DD configurations, and the inverted gap at $\Gamma$ increases during the polarization reversal process from UU to DD (gauged by the out-of-plane electrical dipole $\mu_{\rm OP}$), consistent with the design principle. { It is noted that we carefully check the band energies for bilayers of DU, UD, and DD configurations with a high $k$-point density that samples the whole 2D Brillouin zone (Fig.\,S16 in Supplemental Materials). The values of band gaps in these three configurations, albeit small, are physical. Moreover, the bilayer system with out-of-plane polarization investigated here does not have any crystalline symmetry (e.g., mirror plane) to protect band crossing in the presence of SOC~\cite{Fang15p081201,Gao19p153}.} Calculations of the $\mathbb{Z}_2$ invariant using the Wilson loop approach confirm that both DU and UD configurations are in the QSH insulator phase. As a consequence, topological edge states occur at the boundaries of these samples, which can be clearly seen from the boundary spectral functions shown in Fig.\,\ref{UUDD}f-g.  

Surprisingly, the DD configuration appears to be $\mathbb{Z}_2$ trivial by the same calculation method. This indicates an interesting scenario where the enhanced band inversion can convert a topological insulator to a trivial one through a phase transition, consistent with the model prediction in Fig.\,\ref{phase}a. More important, this highlights that reversing the polarization of a ferroelectric hetero-bilayer, including its metastable intermediate states, generally enables access to various phase points across a wide range of the topological phase diagram.

Following the same design principle, we find a number of bilayer heterostructures consisted of III$_2$-VI$_3$-type 2D FEs are both topological insulator and ferroelectric, i.e., the DD and UD configurations of Al$_2$Te$_3$/Al$_2$Se$_3$, and the DU and UD configurations of Al$_2$Te$_3$/In$_2$S$_3$. The non-trivial band topologies of these systems are confirmed by $\mathbb{Z}_2$ calculations (see Supplemental Materials). 

We comment on the well-known issue of band gap underestimation in (semi-)local DFT such as generalized gradient approximation used in this work.  Our design principle by itself is quite robust and less sensitive to the DFT model as $P_{\rm OP}$ (and thus $\Delta \Phi)$ in principle can always be readily tuned, either by finding appropriate 2D FEs or by applying external electric/stress fields, to satisfy Eq.\,\ref{req1}. 
If a systematic underestimation of the band gap exists, the true phase diagram shown in Fig.\,\ref{UUDD}a will shift to the right horizontally relative to the DFT-predicted one. This may cause both ``false positive" (predicting a non-trivial insulator being trivial) and ``false negative" (predicting a trivial phase being non-trivial) due to the nature of successive phase transitions. We believe high-throughput calculations based on more accurate (albeit more expensive) DFT methods such as hybrid functionals can lead to promising 2DFETIs for experimental synthesis and characterizations. 

{ For a band inversion process driven by SOC, the inversion strength $\lambda$ is intrinsically limited by the atomic numbers of heavy elements contributing to the states near $E_F$. In comparison, the central quantity of the design principle proposed in this work is $P_{\rm OP}$ and associated $E_d$, which can be continuously tuned.  In experimental realizations, we suggest various factors such as the dielectric constants of substrates, unintentional doping due to chemicals used in the device fabrication process, and in-plane strains induced by the lattice mismatch can serve as knobs to obtain precise control of $E_d$ and $\lambda$. For example, one can use another layer of semiconducting 2D material or a substrate to partially screen the surface charges of the bilayer, thus setting the magnitude of $E_d$ and $\lambda$ to the desired value. Surface doping can be utilized to change the surface work function and electron affinity to configure the topological state. }

\begin{figure}[t]
\centering
\includegraphics[scale=1.0]{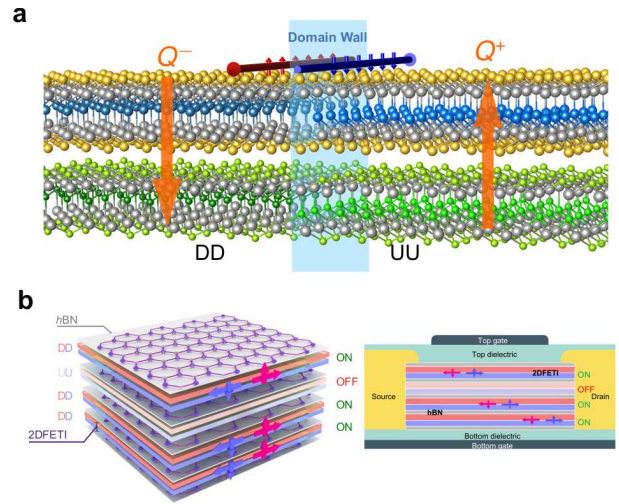}
 \caption{ Schematic diagram of 2DFETI-based devices. A hetero-bilayer is topologically trivial in the UU configuration and non-trivial in the DD configuration. ({\bf a}) Ferroelectric domain walls as field-configurable and moveable quasi-1D channel carrying dissipationless spin current. ({\bf b}) Non-volatile topological memristor made of heterobilayers. The UU configuration is trivial and the DD configuration is nontrivial. The edge conductance can be written to any value between 0 and $2Ne^2/h$ for a stacked structure containing $N$ layers of 2DFETIs. The edge states remain nonvolatile in the absence of vertical electric field between top and bottom gates.}
  \label{device}
 \end{figure}
 
The 2DFETIs exhibit a few features distinct from their 3D counterparts. First, a 2DFETI is expected to possess more robust switchability than a 3DFETI. The conducting surface states of a 3DFETI, though in principle can serve as innate metallic electrodes to stabilize ferroelectricity at the nanoscale~\cite{Liu16p1663}, may also screen strongly the external electric field, hindering the polarization reversal process. In contrast, a 2DFETI behaves like a normal insulator along the out-of-plane direction, making it easier to switch $P_{\rm OP}$. Because the band topology is strongly coupled with the direction and magnitude of $P_{\rm OP}$, it is feasible to use an external electric field to drive a trivial-nontrivial topological phase transition, corresponding to an OFF-ON switch of the quantized edge conductance. According to our design principle (Eq.\,\ref{ti2ni}), the UU and DD configurations of a bilayer heterostructure can have different band topologies. Unlike 1T'-MoS$_2$ that requires a sustained electric field to maintain the trivial state~\cite{Qian14p1344}, the UU and DD configurations are intrinsically stable in the absence of an electric field, allowing non-volatile topological field-effect transistor. Additionally, the 180$^{\circ}$ domain wall (DW) separating UU and DD domains will support helical metallic states protected from back-scattering. These DWs in 2DFETI can serve as field-configurable and moveable quasi-1D dissipationless charge/spin transport channels (Fig.\,\ref{device}a), offering new opportunities of domain-wall-based quantum electrical circuits. 

We propose a topological memristor (illustrated in Fig.\,\ref{device}b) for non-volatile multi-state applications based on vdW heterostructures of 2DFETIs separated by 2D wide-band-gap insulators such as hexagonal boron nitride (hBN). { The device setup is similar to a topological transistor~\cite{Qian14p1344} but with advantages of being nonvolatile and multistate. Noted that a field-effect transistor made completely from 2D materials has already been demonstrated~\cite{Roy14p6259}, indicating the possibility of constructing a similar unit using only 2D materials.} The hetero-bilayer-based 2DFETI is topologically trivial in the UU configuration while non-trivial in the DD configuration. It has been demonstrated in ferroelectric thin films that the polarization state can be deterministically set to a desired value in an on-demand fashion by controlling the voltage and width of pulsed electric fields~\cite{Chanthbouala12p860,Xu19p1282}. Following a similar spirit, the edge conductance of the vdW heterostructure containing $N$ sheets of 2DFETIs can be written electrically to any value between 0 and $2Ne^2/h$ by varying the number of bilayers of DD configurations, and retain the value without bias. The proposed 2DFETI-based topological memristor, combining the advantages of topological insulator, ferroelectrics, and two-dimensional materials, may hold the promise for energy-efficient, high-density synaptic electronics and neuromorphic systems.

\begin{acknowledgments}
J.H., X.D., and S.L. acknowledge the supports from Westlake Education Foundation, Westlake Multidisciplinary Research Initiative Center, and National Natural Science Foundation of China (52002335). Y.K. acknowledges the support from the NRF Grant (2020R1F1A106926111) The computational resource is provided by Westlake HPC Center and the Korea Institute of Science and Technology Information (KISTI).
\end{acknowledgments}
\bibliography{SL}
\end{document}